\documentclass[aps,prb,twocolumn,superscriptaddress,10pt,nofootinbib]{revtex4-2}
\setcitestyle{numbers,square}
\pdfoutput=1

\usepackage[utf8]{inputenc}
\usepackage[T1]{fontenc}
\usepackage[english]{babel} 
\usepackage{amsmath, amsfonts, amssymb}
\usepackage[locale=US, separate-uncertainty=true, range-units=single, list-units=single, retain-unity-mantissa=false]{siunitx}
\usepackage{xspace}
\usepackage{nicefrac}
\usepackage{multirow}
\usepackage{graphicx}
\usepackage{mathtools}
\usepackage{enumerate}
\usepackage{bm}
\usepackage[dvipsnames]{xcolor}
\usepackage{hyperref}
\hypersetup{
	colorlinks = true,
	linkcolor = blue,
	citecolor = blue,
	filecolor = blue,
	urlcolor = blue
}

\usepackage{prettyref}%
\newcommand{\pref}[1]{\prettyref{#1}}%
\newrefformat{fig}{Fig.~\ref{#1}}%
\newrefformat{tab}{Table~\ref{#1}}%
\newrefformat{sec}{Sec.~\ref{#1}}%
\newrefformat{app}{App.~\ref{#1}}%
\newrefformat{eq}{Eq.~(\ref{#1})}%

\DeclareSIUnit\angstrom{\protect \text {Å}}

\newcommand{\BM}{\ensuremath{R_1^+}\xspace}
\newcommand{\JT}{\ensuremath{M_2^+}\xspace}
\newcommand{\eg}{\ensuremath{{e_g}}\xspace}
\newcommand{\ttg}{\ensuremath{{t_{2g}}}\xspace}
\newcommand{\CFO}{CaFeO$_3$\xspace}
\newcommand{\Pton}{\ensuremath{P2_1/n}\xspace}

\begin{document}

\title{Calculation of screened Coulomb interaction parameters for the charge-disproportionated insulator CaFeO$_3$}

\author{Maximilian E. Merkel}
\email{maximilian.merkel@mat.ethz.ch}
\affiliation{Materials Theory, ETH Z\"u{}rich, Wolfgang-Pauli-Strasse 27, 8093 Z\"u{}rich, Switzerland}
\author{Claude Ederer}
\email{claude.ederer@mat.ethz.ch}
\affiliation{Materials Theory, ETH Z\"u{}rich, Wolfgang-Pauli-Strasse 27, 8093 Z\"u{}rich, Switzerland}

\date{\today}

\begin{abstract}
We calculate the screened electron-electron interaction for the charge-disproportionated insulator CaFeO$_3$ using the constrained random-phase approximation (cRPA). While in many correlated materials, the formation of a Mott-insulating state is driven by a large local Coulomb repulsion, represented by the Hubbard $U$, several cases have been identified more recently where $U$ is strongly screened and instead the Hund's interaction $J$ dominates the physics. Our results confirm a strong screening of the local Coulomb repulsion $U$ in CaFeO$_3$ whereas $J$ is much less screened and can thus stabilize a charge-disproportionated insulating state. This is consistent with the case of the rare-earth nickelates where similar behavior has been demonstrated. In addition, we validate some common assumptions used for parametrizing the local electron-electron interaction in first-principles calculations based on density-functional theory (DFT), assess the dependence of the interaction on the choice of correlated orbitals, and discuss the use of the calculated interaction parameters in DFT+$U$ calculations of CaFeO$_3$. Our work also highlights certain limitations for the direct use of cRPA results in DFT-based first-principles calculations, in particular for systems with strong entanglement between the correlated and uncorrelated bands.
\end{abstract}

\maketitle

\section{Introduction}

Strongly correlated materials exhibit many interesting physical phenomena, such as high-temperature superconductivity, colossal magnetoresistance, and metal-insulator transitions \cite{dagotto_correlated_1994, dagotto_colossal_2001, imada_metal-insulator_1998}, which makes them also very attractive candidates for a variety of technological applications, such as, e.g., Mott-transistors~\cite{newns_mott_1998, yang_oxide_2011, mannhart_put_2012, vitale_steep-slope_2017}. Here, the idea is to exploit metal-insulator transitions for achieving higher carrier densities, larger on-off ratios, lower switching voltage, and faster switching times.
The emergence of a metal-insulator transition in strongly correlated materials is typically associated with a large on-site Coulomb repulsion, represented by the Hubbard parameter $U$, which forces the electrons to localize and thereby produces an insulating state. A metal-insulator transition can then be triggered by variations in temperature, strain, doping, applied voltage, etc. 

\begin{figure}
    \centering
    \includegraphics[width=.9\linewidth]{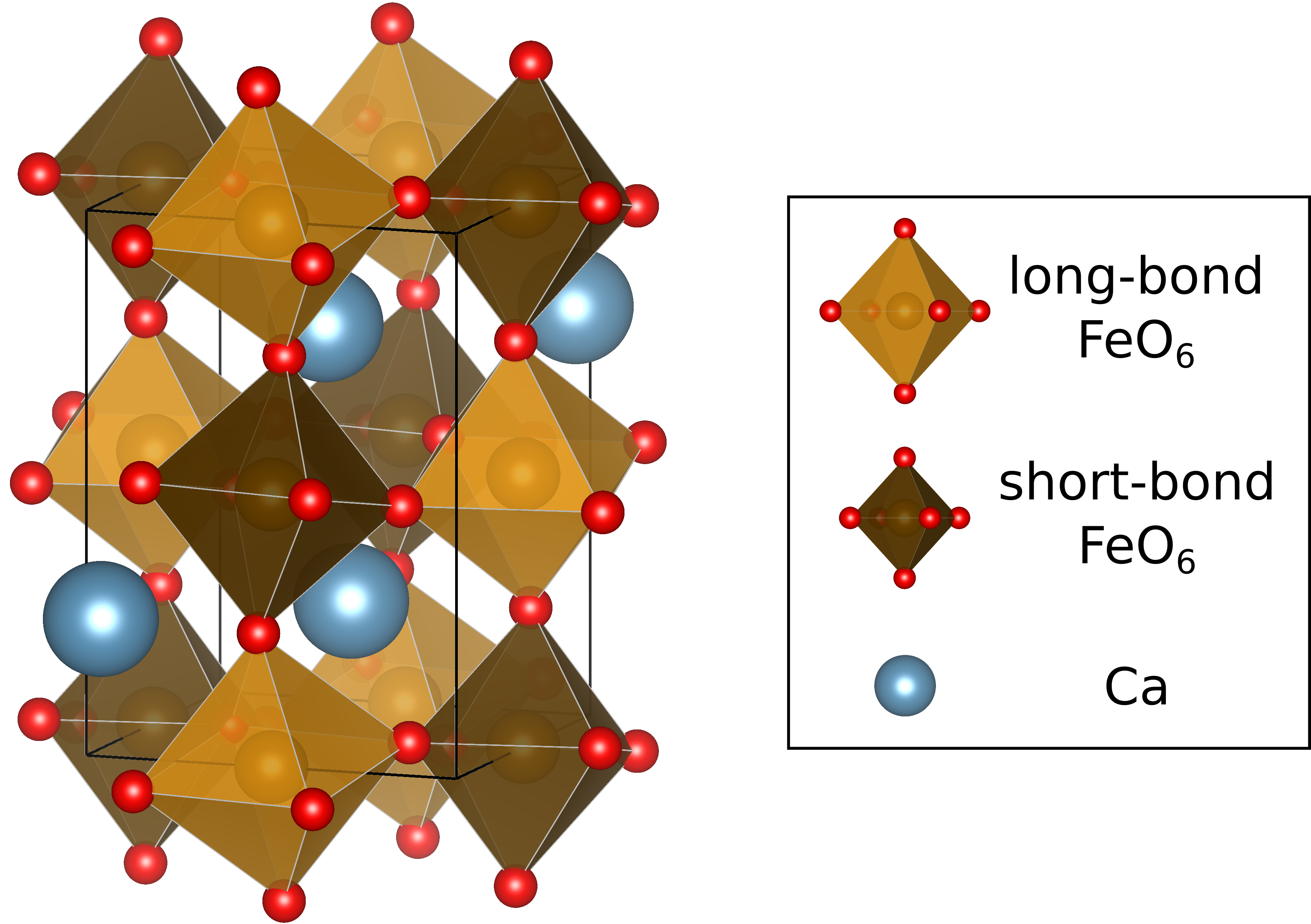}
    \caption{Experimental crystal structure of \CFO below the metal-insulator transition \cite{woodward_structural_2000}. The breathing distortion creates long-bond (light orange) and short-bond (dark orange) FeO$_6$ octahedra, which are rotated and tilted by other distortions.}
    \label{fig:crystal_struct}
\end{figure}

However, for certain transition-metal oxides exhibiting metal-insulator transitions, it was suggested that their basic physics can be explained within a minimal model where the Hubbard $U$ is in fact strongly screened and therefore rather small, and instead the metal-insulator transition is controlled by the strength of the Hund's interaction $J$ \cite{mazin_charge_2007, subedi_low-energy_2015, peil_mechanism_2019, isidori_charge_2019, merkel_charge_2021}.
In these cases, the metal-insulator transition is accompanied by a charge disproportionation of the transition-metal cations, resulting in crystallographically inequivalent sites and a breathing distortion of the surrounding oxygen polyhedra (see Fig.~\ref{fig:crystal_struct}).

Such strong screening of the Coulomb repulsion between the transition-metal $d$ electrons can be caused by the O-$p$ electrons and is the more efficient the closer in energy the O-$p$ states are to the correlated $d$ states. This makes materials with a small or even negative charge-transfer energy good candidates for this type of behavior \cite{mazin_charge_2007,khomskii_transition_2014}, such as oxides containing late transition-metal cations with high oxidation states. Indeed, rare-earth nickelates~\cite{alonso_charge_1999, alonso_room-temperature_2000, catalano_rare-earth_2018} and certain ferrites~\cite{takano_charge_1977, battle_study_1988, guo_21_2017} have been identified as \emph{charge-disproportionated insulators}.
For the case of the well-studied rare-earth nickelates, $R$NiO$_3$, the strong screening of the local Coulomb repulsion was confirmed by {\it ab initio} calculations of the screened interaction parameters~\cite{seth_renormalization_2017, hampel_energetics_2019}.

Long before the relevance of charge disproportionation for the metal-insulator transition in the rare-earth nickelates has been recognized, a metal-insulator transition around room temperature involving charge disproportionation was reported for paramagnetic \CFO \cite{takano_charge_1977,kawasaki_phase_1998}. 
Here, the Fe cation has a nominal average oxidation state of $4+$ and disproportionates according to  $2d^4 \rightarrow d^3 + d^5$, with a high-spin $d^5$ configuration at ambient pressure~\cite{takano_pressure-induced_1991}.
The charge disproportionation in \CFO also couples to a breathing distortion of the FeO$_6$ octahedra~\cite{morimoto_structure_1996, takeda_metalsemiconductor_2000, woodward_structural_2000}, resulting in a symmetry lowering from $Pbnm$ to \Pton and a transition from metal to insulator \cite{kawasaki_phase_1998, rogge_electronic_2018}. The low-temperature \Pton crystal structure of \CFO is depicted in \pref{fig:crystal_struct}.

Several computational studies based on density-functional theory (DFT) have addressed the charge disproportionation and metal-insulator transition in \CFO in order to clarify the underlying physics~\cite{whangbo_effect_2002, saha-dasgupta_density_2005, sadoc_role_2007, alexandrov_ab_2008, cammarata_spin-assisted_2012, wang_quantifying_2014, dalpian_bond_2018, zhang_charge-_2018, leonov_metal-insulator_2022}. Thereby, to obtain a realistic description of the insulating state, the local electron-electron interaction is typically explicitly treated within DFT+$U$~\cite{anisimov_first-principles_1997} or DFT+dynamical mean-field theory (DMFT)~\cite{georges_dynamical_1996,held_electronic_2007}.
Different empirically chosen interaction parameters have been used in these calculations, ranging from $U = \SI{3}{eV}$ to more than \SI{7}{eV} for the Hubbard parameter, and from $J = \SI{0}{eV}$ to as much as \SI{2}{eV} for the Hund's interaction. A first-principles-based calculation of screened interaction parameters using cRPA can also give insights on whether these choices are physically reasonable.
In our recent tight-binding+DMFT study \cite{merkel_charge_2021}, we have found a very rich phase diagram for paramagnetic \CFO as a function of $U$ and $J$, which shows that using suitable interaction parameters is crucial for obtaining the correct phase. 
However, we also note that different studies often use different definitions of the correlated local Fe-$d$ orbitals, which hampers a systematic comparison of the corresponding interaction parameters.

Here, we calculate screened interaction parameters for \CFO using the constrained random-phase approximation (cRPA)~\cite{aryasetiawan_frequency-dependent_2004, miyake_ab_2009, sasioglu_effective_2011}.
Our results confirm a strong screening of the Hubbard parameter $U$, while the Hund's interaction $J$ is less affected by the screening. We also compare different choices of the correlated $d$ orbitals, corresponding either to an atomic-orbital-like basis, as commonly used in DFT+$U$, or to more molecular-orbital-like hybridized Fe-$d$/O-$p$ frontier orbitals, which closely resemble the nominal oxidation states and are more commonly used in model Hamiltonian or DFT+DMFT studies. Furthermore, we assess the validity of the typically-used spherical approximation for the local electron-electron interaction and discuss the applicability of the calculated screened interaction parameters in DFT+$U$ and DFT+DMFT calculations of \CFO.
To that end, we also revisit the $U$ (and $J$) dependence of the relaxed high- and low-temperature structure of \CFO within DFT+$U$ and compare the interaction parameters used for these structural relaxations with the ones obtained within cRPA.

\section{Methods}

In this section, we introduce the interaction Hamiltonian acting on the correlated subspace of Fe-$d$ electrons and discuss the determination of the corresponding screened interaction parameters using cRPA. Then, we describe the details and all parameters used in our DFT(+$U$) calculations and in the construction of the Wannier basis representing the correlated orbitals.

\subsection{Interaction Hamiltonian and constrained random-phase approximation} \label{sec:methods_h_int}

The general local two-particle interaction Hamiltonian can be written as follows (see, e.g., \cite{pavarini_ldadmft_2011}):
\begin{align} \label{eq:h_int_full}
    H_\mathrm{int} = \frac12 \sum_{\substack{m m' m'' m''' \\ \sigma \sigma'}} U_{mm'm''m'''} c^\dag_{m \sigma} c^\dag_{m' \sigma'} c_{m''' \sigma'} c_{m'' \sigma} , 
\end{align}
where $c_{m \sigma}$ is the annihilation operator of an electron with spin $\sigma$ in orbital $m$ and the matrix elements $U_{m m' m'' m'''}$ are defined through
\begin{align}
U_{m m' m'' m'''} = \int d \bm r d \bm r' [&\phi_{m}^*(\bm r) \phi_{m'}^*(\bm r') v(\bm r, \bm r') \nonumber\\
&\times\phi_{m'''}(\bm r') \phi_{m''}(\bm r) ] .
\label{eq:U-4-index}
\end{align}
Here, $\phi_m(\bm{r})$ are the orbitals and $v(\bm{r},\bm{r}')$ is the interaction between these orbitals, which in the following can be either the bare or the screened Coulomb interaction.

For practical calculations, the full four-index form of $U_{mm'm''m'''}$ is sometimes reduced to a two-index form according to $U_{mm'} = U_{mm'mm'}$ and $J_{mm'} = U_{mm'm'm}$. These two-index terms are assumed to dominate the interaction.
The third possible two-index matrix $I_{mm'} = U_{mmm'm'}$ equals $J_{mm'}$ if the orbitals $\phi_m(\bm r)$ are real-valued, which can be directly seen from \pref{eq:U-4-index}. In the following, we always use real-valued, maximally localized Wannier functions~\cite{marzari_maximally_1997}, and thus we only discuss $U_{mm'}$ and $J_{mm'}$. 

The interpretation of these two-index terms becomes clear when considering only the density-density terms of the Hamiltonian \begin{align} \label{eq:h_int}
    H_\mathrm{int} = &\frac12 \sum_{mm',\sigma} U_{mm'} n_{m\sigma} n_{m'\bar\sigma} \nonumber\\
    + &\frac12 \sum_{m\neq m',\sigma} (U_{mm'} - J_{mm'}) n_{m\sigma} n_{m'\sigma},
\end{align}
with the density operators $n_{m\sigma} = c_{m\sigma}^\dag c_{m\sigma}$ and the convention $\bar\sigma = -\sigma$.
Here, $U_{mm'} = U^{\sigma\bar\sigma}_{mm'}$ represents the interaction between electrons with opposite spins and $U_{mm'} - J_{mm'} = U^{\sigma\sigma}_{mm'}$ the interaction between electrons with the same spin.

To obtain screened interaction parameters within cRPA \cite{aryasetiawan_frequency-dependent_2004, miyake_ab_2009, sasioglu_effective_2011, kaltak_merging_2015}, the Hilbert space is divided into the \emph{correlated subspace}, represented by the orbitals $\phi_m({\bf r})$ for which $H_\mathrm{int}$ is defined, and the rest.
Then, the bare electron-electron interaction $v(\bm r, \bm r') = |\bm r - \bm r'|^{-1}$ is screened by the polarization function $P_\mathrm{rest}$ including all electronic transitions that involve the rest of the Hilbert space, but not those purely within the correlated subspace itself. The screened Coulomb interaction $U$ is then given by $U = [1-v P_\mathrm{rest}]^{-1} v$ \cite{aryasetiawan_frequency-dependent_2004}.

Typically, $P_\mathrm{rest}$ is calculated as $P_\mathrm{total} - P_\mathrm{corr}$, where $P_\mathrm{total}$ is the total polarization, which can easily be calculated from the Kohn-Sham wave-functions, and $P_\mathrm{corr}$ is the polarization only within the correlated subspace.
If the correlated bands are not entangled with the rest, $P_\mathrm{corr}$ is well-defined and can also be calculated from the Kohn-Sham wave-functions of the corresponding bands.
If there is entanglement between the correlated and uncorrelated subspace, different approaches have been introduced to calculate the polarization function.

One option is to neglect the hybridization between the correlated subspace and the rest subspace in the Hamiltonian, and diagonalize these two subspaces individually. This leads to two independent sets of wave functions and eigenenergies, which for small hybridization differ only slightly from the original Kohn-Sham bands. The polarizations $P_\mathrm{total}$ and $P_\mathrm{corr}$ can then be calculated from these new wave functions and eigenenergies \cite{miyake_ab_2009}. This method is called the ``disentanglement method'' in the following.

Alternatively, Ref.~\onlinecite{sasioglu_effective_2011} suggested to obtain $P_\mathrm{corr}$ by assigning weights to each transition in the polarization function that measure the probability of an electron residing in the correlated suspace both before and after the transition. These weights are based on the projections of the Bloch functions onto the correlated orbitals. We refer to this method as the ``weighted method''.

Finally, Ref.~\onlinecite{kaltak_merging_2015} proposed a slightly different expression for $P_\text{corr}$, derived from the general Kubo-Nakano formula for the response function corresponding to density fluctuations within the correlated subspace.
We call this the ``projector method'' in the remainder of this work.
It also is the default option implemented in the Vienna Ab-initio Simulation Package (VASP)~\cite{kresse_ab_1993, kresse_efficient_1996} and is used throughout this work, except in~\pref{sec:res_crpa_disent}, where we present an explicit comparison between all three methods for obtaining the polarization function.

\subsection{Spherical parametrization and symmetrization of the interaction}
\label{sec:method_crpa_avg}

For DFT+$U$ and DFT+DMFT calculations, the interaction matrix $U_{mm'm''m'''}$ is typically constructed assuming spherical symmetry, i.e, $v(\bm r, \bm r') = v(|\bm r - \bm r'|)$ and atomic-like orbitals $\phi_m(\bm r)$, where $m$ is the magnetic quantum number corresponding to an electronic shell with well-defined orbital momentum $l$.
This allows to expand the interaction $v$ in terms of complex spherical harmonics $Y_{kq}$ and to arrive at a unique parametrization of the interaction matrix in terms of $l+1$ independent parameters, the Slater integrals $F_k$, where $k$ is an even integer with $0 \leq k \leq 2l$ (see, e.g., \cite{pavarini_ldadmft_2011, vaugier_hubbard_2012, strand_correlated_2013}).
The resulting interaction matrix in the basis of complex spherical harmonics is
\begin{align}
    U_{mm'm''m'''} = \sum_{k=0}^{2l} F_k \alpha^{(k)}_{mm'm''m'''}
    \label{eq:def_u_slater}
\end{align}
with
\begin{align}
    \alpha^{(k)}_{mm'm''m'''} =& \frac{4\pi}{2k+1} \sum_{q=-k}^k \langle lm | Y^*_{kq} | lm'' \rangle \langle lm' | Y_{kq} | lm''' \rangle \nonumber\\
    =& (2l+1)^2 \begin{pmatrix} l & k & l \\ 0 & 0 & 0 \end{pmatrix}^2 (-1)^{m+m'''} \nonumber\\
    &\times\begin{pmatrix} l & k & l \\ -m & m-m'' & m'' \end{pmatrix} \nonumber\\
    &\times \begin{pmatrix} l & k & l \\ -m' & m'-m''' & m''' \end{pmatrix} \delta_{m''-m, m'-m'''}
    \label{eq:def_alpha_3j}
\end{align}
where the ``$2\times3$ matrices'' are Wigner 3-$j$ symbols. 

The Slater integrals are in principle defined as the radial integrals appearing in this expansion. Conversely, they can be extracted from the interaction matrix by inverting \pref{eq:def_u_slater} and using the orthogonality relation of the $\alpha$ tensor as specified in \pref{eq:ortho_alpha}:
\begin{align}
    F_k &= c_{lk}^{-1} \sum_{m m' m'' m'''} U_{m m' m'' m'''} \alpha^{(k)}_{mm'm''m'''} \nonumber\\[10pt] 
    c_{lk} &= \frac{(2l+1)^4}{2k+1} \begin{pmatrix} l & k & l \\ 0 & 0 & 0 \end{pmatrix}^4
    \label{eq:avg_umatrix_slater}
\end{align}
\pref{eq:avg_umatrix_slater} allows us to reduce any interaction matrix, e.g., one obtained from cRPA to a set of Slater parameters, which can then be used in \pref{eq:def_u_slater} to construct a spherically symmetrized interaction matrix%
    \footnote{Note that our definition of the Slater integrals in \pref{eq:avg_umatrix_slater} differs from Eq.~(B3) of Ref.~\onlinecite{vaugier_hubbard_2012} due to the appearance of the Kronecker $\delta_{m''-m, m'-m'''}$ in \pref{eq:def_alpha_3j}. Both forms are equivalent if the matrix $U_{m m' m'' m'''}$ is already spherically symmetric, but for a general interaction, e.g., one obtained from cRPA, only \pref{eq:avg_umatrix_slater} corresponds to the correct inversion of \pref{eq:def_u_slater}.}.
Note that while these formulas are specified in the basis of complex spherical harmonics, in \pref{sec:dft_results} we represent the interaction matrices in the basis of real cubic harmonics. These two types of functions are related by a simple linear transformation.

Often, the parametrization of the interaction is further simplified to only two parameters $U$ and $J$, which are defined in terms of the two-index matrices:
\begin{align}
    U &= \frac1{(2l+1)^2} \sum_{m m'} U_{m m'}, \nonumber\\
    U - J &= \frac1{2l(2l+1)} \sum_{m m'} (U_{m m'} - J_{m m'})
    \label{eq:avg_umatrix}
\end{align}
From \pref{eq:avg_umatrix_slater} and \pref{eq:avg_umatrix}, it follows directly that $F_0 \equiv U$ and, for $l=2$, $J \equiv (F_2 + F_4)/14$. (The general relation between Slater integrals and $J$ for all $l$ is given in \pref{eq:general_j_slater}).
However, the inverse operation of obtaining the interaction matrix from only $U$ and $J$ is underdefined for $l \geq 2$ since this requires knowledge of all $l+1$ Slater integrals. Therefore, a fixed ratio of $F_4/F_2 \approx 0.63$  is usually assumed for $d$ electrons~(see, e.g.,~\cite{pavarini_ldadmft_2011}), which is based on calculations of the unscreened interaction using atomic-orbital-like orbitals \cite{de_groot_2p_1990, anisimov_density-functional_1993}.

Within cRPA, it is straightforward to not only calculate the local elements of the interaction matrix
between orbitals on the same site but also between orbitals on different sites in the same unit cell. 
For practical purposes, it is often assumed that the intersite terms are not strongly orbitally dependent \cite{campo_extended_2010}. Therefore, we only discuss the orbitally averaged nearest-neighbor (NN) interaction
\begin{align}
    V_\mathrm{NN} &= \frac1{(2l+1)^2} \sum_{m m'} U_{m m'}, \nonumber\\
    J_\mathrm{NN} &= \frac1{(2l+1)^2} \sum_{m m'} J_{m m'} ,
    \label{eq:avg_umatrix_nn}
\end{align}
where now $m$ and $m'$ are on NN sites.

\subsection{DFT calculations and Wannier orbitals} \label{sec:methods_dft}

For all calculations, we use the VASP code (version 6.3.0) \cite{kresse_ab_1993, kresse_efficient_1996} in combination with the PBEsol exchange-correlation functional \cite{perdew_restoring_2008}. We use standard projector augmented wave (PAW) potentials \cite{blochl_projector_1994, kresse_ultrasoft_1999} with valence states 3$s^2$3$p^6$4$s^2$ for Ca, 3$s^2$3$p^6$3$d^7$4$s^1$ for Fe, and 2$s^2$2$p^4$ for O. We treat \CFO in a 20-atom unit cell, which allows us to describe both the $Pbnm$ and \Pton structures, using a $7 \times 7 \times 5$ k-point mesh, a plane-wave energy cutoff of \SI{600}{eV}, and an energy tolerance of \SI{e-8}{eV} between electronic steps, which leads to good convergence.

Structural relaxations are performed up to an energy tolerance of \SI{e-7}{eV} between ionic steps. Except where explicitly noted, relaxations of the unit cell and atomic positions are done using DFT+$U$ \cite{liechtenstein_density-functional_1995} with $U = \SI{4}{eV}$ and $J = \SI{1}{eV}$, similar to previous studies \cite{saha-dasgupta_density_2005, cammarata_spin-assisted_2012, dalpian_bond_2018}, but we also present results where we systematically vary $U$ over a wider range.
For the structural relaxation, we emulate the complex low-temperature helical magnetic order in \CFO \cite{woodward_structural_2000, rogge_itinerancy-dependent_2019} with the much simpler A-type antiferromagnetic ordering, which has the lowest energy out of A-, C- and G-type orderings in our calculations.
For the high-temperature $Pbnm$ structure, we also perform calculations where we relax only the atomic positions using spin-degenerate DFT without +$U$ correction, which we call ``bare DFT'' here.

We compare the relaxed to the experimental structures using the mode-decomposition tool ISODISTORT \cite{stokes_isodistort_2022, campbell_isodisplace_2006}. All amplitudes are normalized to the 20-atom unit cell, and the symmetry-labeling convention uses a cubic perovskite parent cell with Fe at the origin.
The most important modes are the \BM breathing mode in the \Pton structure, which splits the Fe into two inequivalent sites, and the \JT Jahn-Teller mode in the $Pbnm$ structure, which locally splits the Fe-\eg orbitals but experimentally is not relevant in \CFO. Also present in $Pbnm$ are the modes related to octahedral rotations and tilting, $M_3^+$, $R_4^+$, and $X_5^+$, which mainly decrease the hopping and therefore the band width~\cite{saha-dasgupta_density_2005}. Additionally, there are three more symmetry-allowed modes with very small amplitudes: $R_5^+$ ($Pbnm$), $R_3^+$, and $M_5^+$ (both \Pton).

We use Wannier90 (version 3.1.0) \cite{pizzi_wannier90_2020} to construct the Wannier orbitals defining the correlated subspace for our cRPA calculations.
We start with initial projections corresponding to $d$-shell cubic harmonics defined in a local coordinate system aligned along the Fe-Fe distances.
Thereby, we employ a disentanglement (outer) window from \SIrange{-10}{5}{eV} relative to the Fermi energy, which is chosen to be large enough to encompass all 56 O-$p$ and Fe-$d$ derived bands, and, optionally, a frozen (inner) window from \SIrange{-1.1}{5}{eV}.
We always first perform the disentanglement, where convergence is defined as a relative change of less than \num{e-9} of the gauge-invariant part of the spread, and then the localization (Wannierization) down to an absolute change of less than \SI{e-9}{\angstrom^2} of the total spread.
Note that the resulting Wannier functions still resemble the cubic harmonics, which is required to map the cRPA results onto \pref{eq:avg_umatrix_slater}.

For the cRPA calculations, we first converge a nonmagnetic DFT calculation with 512 bands and otherwise the same parameters as listed above. Then we perform one exact-diagonalization step in DFT to obtain the Kohn-Sham band energies.
We perform cRPA calculations only for the zero frequency component of the screened interaction and with the correlated subspace defined by the Wannier orbitals.

\section{Results}\label{sec:dft_results}

\subsection{Structural relaxations} \label{sec:res_dft_relax}

\begin{table*}
    \caption{Lattice parameters and distortion mode amplitudes (with respect to the cubic $Pm\bar{3}m$ high symmetry reference structure) obtained from DFT+$U$ structural relaxations within $Pbnm$ and $P2_1/n$ symmetries. The third line corresponds to a calculation where lattice parameters are fixed to those obtained from DFT+$U$ and only the atomic positions are relaxed using spin-degenerate DFT without a ``$+U$'' correction. The last two lines correspond to experimental data from Ref.~\onlinecite{woodward_structural_2000} obtained at 15\,K ($P2_1/n$) and 300\,K ($Pbnm$).
    The $M_5^+$, \BM and $R_3^+$ modes are not allowed in $Pbnm$ symmetry.
    }
    \label{tab:dft_relax}
    \centering
    \begin{ruledtabular}
    \begin{tabular}{l|cccc|ccccc|ccc}
        & \multicolumn{4}{c|}{Unit cell} & \multicolumn{7}{c}{Mode amplitudes (\si{\angstrom})} \\
        & $a$ (\si{\angstrom}) & $b$ (\si{\angstrom}) & $c$ (\si{\angstrom}) & $\beta$ (\si{\degree}) & \JT & $M_3^+$ & $R_4^+$ & $R_5^+$ & $X_5^+$ & $M_5^+$ & \BM & $R_3^+$ \\
        \hline
        DFT+$U$ \Pton & 5.29 & 5.33 & 7.43 & 89.85 & 0.01 & 0.75 & 1.12 & 0.09 & 0.45 & 0.01 & 0.16 & 0.05 \\
        DFT+$U$ $Pbnm$ & 5.27 & 5.41 & 7.38 & 90.00 & 0.24 & 0.79 & 1.13 & 0.10 & 0.52 &&& \\
        bare DFT $Pbnm$ & \multicolumn{4}{c|}{\textit{fixed to DFT+$U$ $Pbnm$ results}} & 0.03 & 0.89 & 1.10 & 0.12 & 0.53 &&& \\
        Exp. \Pton \cite{woodward_structural_2000} & 5.31 & 5.35 & 7.52 & 90.07 & 0.04 & 0.82 & 1.05 & 0.11 & 0.40 & 0.01 & 0.18 & 0.05 \\
        Exp. $Pbnm$ \cite{woodward_structural_2000} & 5.33 & 5.35 & 7.54 & 90.00 & 0.01 & 0.77 & 0.99 & 0.05 & 0.36 &&&
    \end{tabular}
    \end{ruledtabular}
\end{table*}

We first perform full structural relaxations of CaFeO$_3$ within both $Pbnm$ and $P2_1/n$ symmetries using DFT+$U$ and A-type antiferromagnetic order, as described in Sec.~\ref{sec:methods_dft}. The corresponding results are listed in Table~\ref{tab:dft_relax}, together with experimental data from Ref.~\onlinecite{woodward_structural_2000}.

It can be seen that the lattice parameters obtained for the $Pbnm$ and $P2_1/n$ structures differ only very little, with only a small deviation from $\beta=\ang{90}$ in the monoclinic case. Furthermore, the agreement with the experimental lattice parameters is very good (there is only a slight underestimation of the $c$ lattice parameter by \SI{2.1}{\percent} and the monoclinic distortion is slightly lower than \ang{90} rather than larger). 

The calculated mode amplitudes also agree very well with the experimental data, except for the \JT Jahn-Teller mode in the $Pbnm$ case, which is essentially zero in the room-temperature structure of Ref.~\onlinecite{woodward_structural_2000}, whereas a noticeable Jahn-Teller distortion develops in the zero temperature DFT+$U$ A-type antiferromagnetic case. This is due to the nominally Jahn-Teller-active $d^4$ electron configuration of the Fe$^{4+}$ cation, and has also been found with a very similar amplitude of \SI{.26}{\angstrom} in a previous DFT+$U$ study of \CFO if the symmetry is constrained to $Pbnm$~\cite{zhang_charge-_2018}.

In the $P2_1/n$ structure, which is about \SI{11}{meV} per formula unit lower in energy than the $Pbnm$ structure, the \JT mode nearly vanishes and instead the \BM breathing mode appears with a magnitude of \SI{.16}{\angstrom}, in very good agreement with the experimental data.
Furthermore, it can be seen that the modes related to octahedral tilts and rotations, $R_4^+$ and $M_3^+$ (and also $X_5^+$), do not change significantly between $Pbnm$ and $P2_1/n$, and also agree well with the experimental data.

\begin{figure}
    \centering
    \includegraphics[width=1\linewidth]{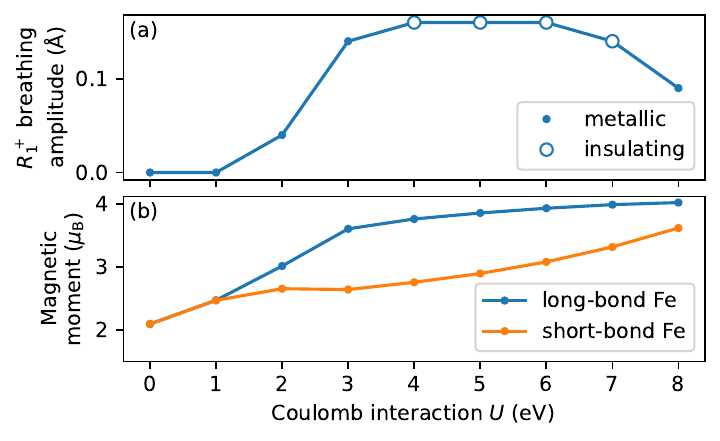}
    \caption{(a) \BM breathing mode amplitude and (b) local magnetic moments of the two inequivalent Fe sites obtained from full structural relaxation within \Pton symmetry as a function of $U$ at $J = \SI{1}{eV}$.
    }
    \label{fig:uscan}
\end{figure}

\pref{fig:uscan} shows the evolution of the relaxed \BM breathing mode amplitude and the local magnetic moments of the two inequivalent Fe sites as function of the Hubbard $U$ parameter for fixed $J=1$\,eV in the \Pton structure.
We note that within the DFT+$U$ method, $U$ and $J$ refer to rather localized atomic-like orbitals (similar to what we refer to as the localized basis in \pref{sec:bands}). This basis does not correspond to the minimal models for charge disproportionation discussed, e.g., in Refs.~\onlinecite{mazin_charge_2007, subedi_low-energy_2015, peil_mechanism_2019, isidori_charge_2019, merkel_charge_2021} where the Hund's interaction drives the transition to the charge-disproportionated insulating state. Instead, as seen in \pref{fig:uscan}, the Hubbard $U$ remains the main control parameter in these DFT+$U$ calculations.
The \BM amplitude becomes nonzero for $U \geq 2$\,eV and is accompanied by a sizeable difference in the local magnetic moments, indicating the disproportionation of the Fe sites into nominal $d^3$ and $d^5$ valences. For $4\,\mathrm{eV} \leq U \leq 7$\,eV the system is insulating and the \BM amplitude stays nearly constant. For larger $U>7$\,eV, the breathing mode amplitude decreases and the system becomes metallic again.
This is similar to what has been observed for the rare-earth nickelates in Ref.~\onlinecite{hampel_interplay_2017}, where a large $U$ shifts the transition-metal $d$ levels below the O-$p$ states in energy, which then disfavors the charge disproportionation. Thus, even though a small or maybe even slightly negative charge-transfer energy seems to be necessary for this type of charge disproportionation, it appears unfavorable if the charge-transfer energy becomes too negative.

Overall, our results show that structural parameters in good agreement with available experimental data can be obtained within DFT+$U$ using interaction parameters of $U = \SI{4}{eV}$ and $J = \SI{1}{eV}$ and an appropriate magnetic order, consistent with previous work~\cite{cammarata_spin-assisted_2012, dalpian_bond_2018}.

Finally, to asses the effect of the imposed A-type magnetic order on the high-temperature $Pbnm$ structure, we perform an additional relaxation for the spin-degenerate case using $U=0$\,eV, where we fix the lattice parameters to the ones obtained from the $Pbnm$ DFT+$U$ relaxation and then relax only the internal atomic positions. We note that fixing the lattice parameters to the DFT+$U$ values is necessary, since the spin-degenerate DFT calculation leads to a low-spin state of the Fe cation which would in turn lead to a significantly lower unit cell volume.
\pref{tab:dft_relax} shows that, in this case, the \JT Jahn-Teller mode disappears almost completely, while all other mode amplitudes are rather similar to the magnetic DFT+$U$ calculation.
This shows that the Jahn-Teller mode is indeed stabilized by the +$U$ correction in combination with the magnetic order, while the octahedral rotation modes are unaffected by this.
Throughout the remainder of this work, we use the so-obtained structure as the high-temperature $Pbnm$ structure unless otherwise noted.

\subsection{Band structure and the correlated subspace}
\label{sec:bands}

\begin{figure}
    \centering
    \includegraphics[width=1\linewidth]{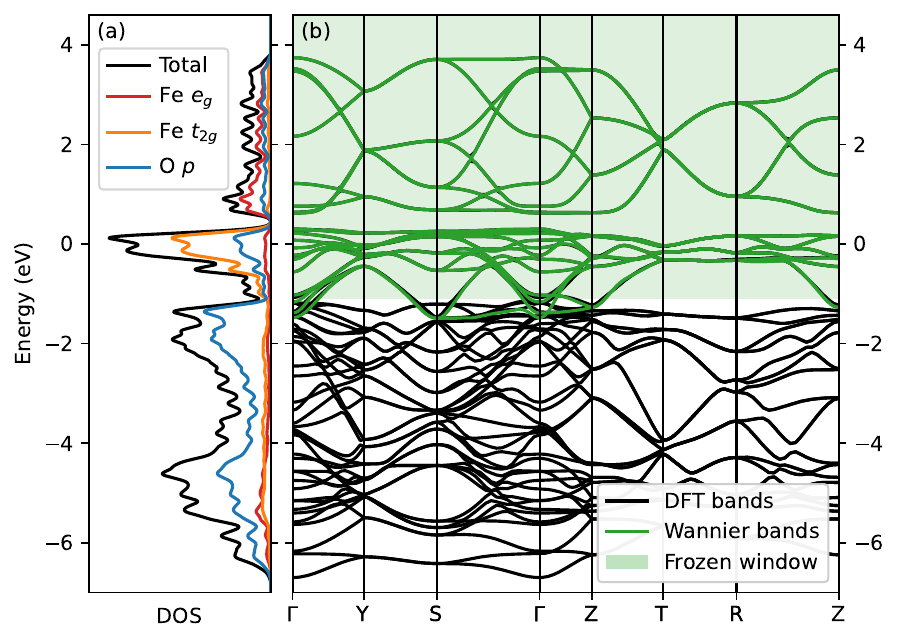}
    \includegraphics[width=1\linewidth]{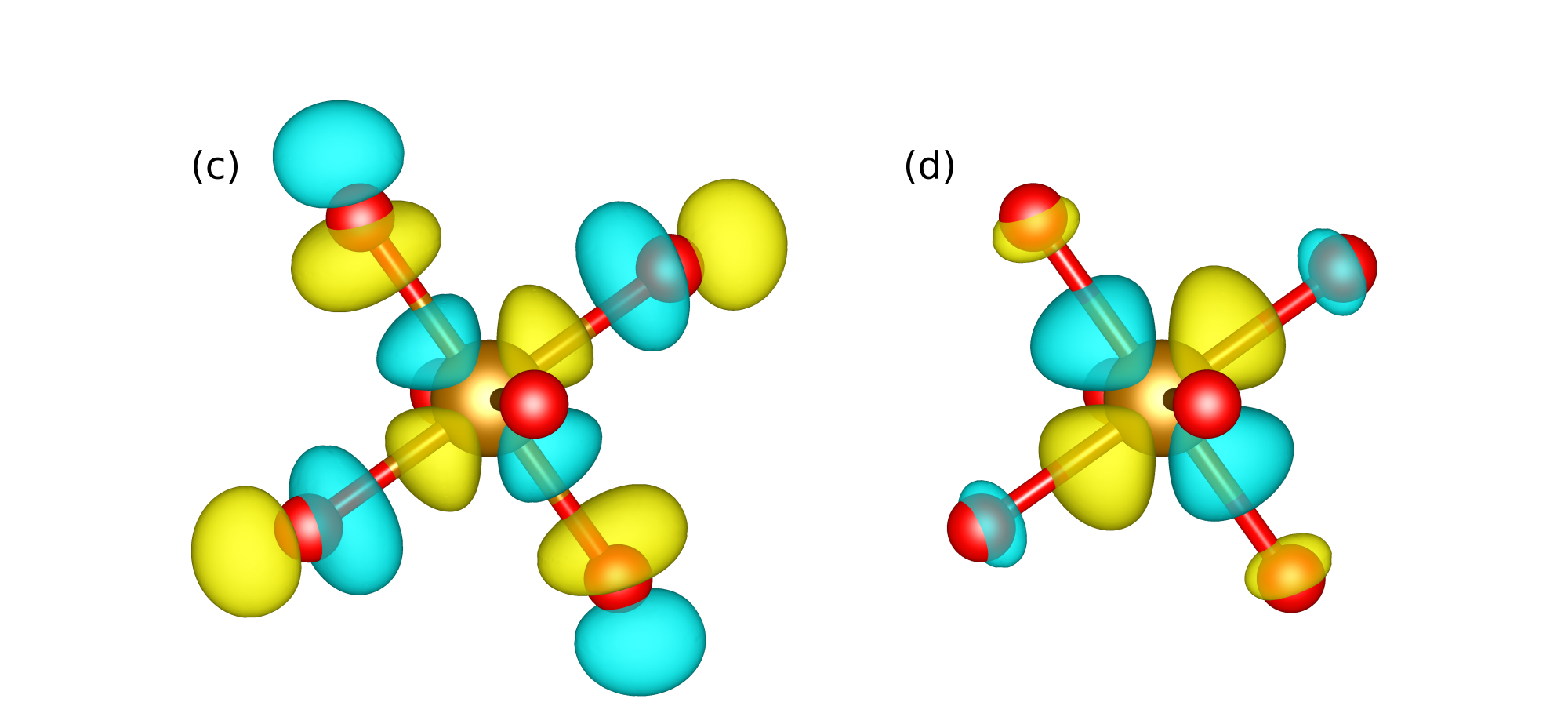}
    \caption{(a) Total and projected density of states (DOS) and (b) band structure along high symmetry lines of the orthorhombic Brillouin zone of \CFO, obtained from the spin-degenerate DFT calculation for the $Pbnm$ structure. The green shaded region in (b) indicates the frozen energy window used to construct the frontier-type Wannier orbitals, which result in the dispersion shown by the green lines.
    (c, d) Isosurfaces of the ($x^2-y^2$)-type Wannier orbital (c) in the frontier basis and (d) in the localized basis. Brown (red) spheres indicate Fe (O) atoms.
    }
    \label{fig:bands}
\end{figure}

Next, we discuss the electronic structure obtained from spin-degenerate DFT calculations without a +$U$ correction in the $Pbnm$ structure. \pref{fig:bands} (a) and (b) show the total density of states (DOS), its projection on different atoms and orbitals, as well as the band structure around the Fermi level.
The DOS shows three (partially overlapping) groups of bands with mixed atomic/orbital character due to hybridization. Bands with predominant Fe-\eg character above \SI{.5}{eV} are separated from lower-lying bands with predominant Fe-\ttg character between about \SIrange{-1}{0.5}{eV} due to a strong octahedral crystal field splitting. The Fe-\ttg dominated bands overlap slightly with O-$p$ dominated bands below \SI{-1}{eV}.
The fact that the Fe-$d$ dominated bands are higher in energy than the O-$p$ dominated ones but also overlap slightly indeed indicates a small positive but nearly vanishing charge-transfer energy within the classification scheme according to Zaanen, Sawatzky, and Allen~\cite{zaanen_band_1985}.

We now construct two different sets of Wannier orbitals, starting from initial projections on Fe-centered $d$ orbitals. 
The first set represents a so-called \emph{frontier basis} that describes only the Fe-$d$ dominated bands in the energy range above approximately $-1$~eV and directly maps on minimal models describing the basic physics of charge disproportionation~\cite{strand_valence-skipping_2014, subedi_low-energy_2015, isidori_charge_2019, merkel_charge_2021}.  The second set corresponds to a \emph{localized basis}, resembling an atomic-orbital-like basis similar to what is typically used to define the correlated orbitals within DFT+$U$ calculations.

In both cases, we use an outer energy window comprising all bands shown in \pref{fig:bands} (a) and (b). For the frontier basis, we use an additional inner (frozen) energy window for the disentanglement procedure which comprises all of the Fe-\eg dominated bands as well as the upper part of the Fe-\ttg dominated bands down to \SI{-1.1}{eV}, as indicated in \pref{fig:bands} (b) by the green shading. We note that since this inner energy window encompasses most of the Fe-$d$ dominated bands, except for the narrow overlapping region immediately below $-1$~eV, the specific choice of the outer energy window does not have a strong influence on the resulting Wannier functions.
The corresponding Wannier bands [green lines \pref{fig:bands} (b)] remain restricted to the energy range of the Fe-$d$ dominated bands, and perfectly reproduce the DFT bands [black lines in \pref{fig:bands} (b)] within the inner energy window. Since this inner energy window contains the Fermi level and there is no entanglement with higher lying bands, the Wannier orbitals recover an occupation of exactly four electrons per Fe site, corresponding to the nominal valence of the Fe$^{4+}$ cation.

The resulting Wannier orbitals are centered on the Fe sites but also exhibit strong ``tails'' on the surrounding oxygen ligands, representing the strongly hybridized character of the corresponding bands. \pref{fig:bands}(c) depicts the ($x^2-y^2$)-type Wannier function as an example. One can recognize the antibonding character of the underlying hybridization between Fe-$d$ and O-$p$ orbitals. Such frontier orbitals are often used in DFT plus dynamical mean-field theory calculations or other more model-based approaches that consider only the effective $d$ bands (see, e.g., \cite{merkel_charge_2021}).

For the construction of the localized basis, we do not use a frozen inner window during the disentanglement.
This results in more localized orbitals that closely resemble atomic orbitals and, as already mentioned, are similar to the orbitals used for the projected DOS shown in \pref{fig:bands}(a) and for the +$U$ correction applied in the VASP code~\cite{blochl_projector_1994}. 
\pref{fig:bands}(d) shows the resulting ($x^2-y^2$)-type Wannier function, which does not exhibit the pronounced $p$-like tails on the surrounding  oxygen ligands (small $s$-like tails remain, due to a corresponding weak hybridization and the orthogonalization between Wannier functions centered on different sites).

We note that the additional structural distortions in the low-temperature \Pton structure do not lead to substantial changes of the band structure around the lower bound of the frozen energy window, so that we can use the same energy window relative to the Fermi energy for both crystal structures. 

\subsection{Screened-interaction parameters from cRPA}

\subsubsection{Spherical symmetrization and effective interaction parameters}
\label{sec:crpa_matrices}

We first discuss the effect of the spherical symmetrization of the interaction parameters by comparing the matrix elements obtained directly from cRPA with their spherically averaged form, i.e., by using \pref{eq:avg_umatrix_slater} and \pref{eq:def_u_slater}. Thereby, we focus on the two-index interaction matrices $U^{\sigma\bar\sigma}_{mm'}$ and $U^{\sigma\sigma}_{mm'}$ between electrons with opposite and same spins, respectively.
We note that the largest interaction that is not contained in the two-index matrices is $U_{yz,z^2,yz,x^2-y^2} = \SI{.19}{eV}$ (and equivalent matrix elements) in the frontier basis, and $U_{yz,z^2,yz,x^2-y^2} = \SI{.29}{eV}$ in the localized basis, which is clearly smaller than the dominant components of the two-index matrices (see below).

In the frontier basis, the two-index matrices obtained directly from cRPA are
\begin{align}
    U_\mathrm{cRPA}^{\sigma\sigma} &= \begin{pmatrix*}[r]
        0.00 & 1.09 & 1.01 & 1.10 & 1.61 \\
        1.09 & 0.00 & 1.47 & 1.12 & 1.15 \\
        1.01 & 1.47 & 0.00 & 1.47 & 1.01 \\
        1.10 & 1.12 & 1.47 & 0.00 & 1.16 \\
        1.61 & 1.15 & 1.01 & 1.16 & 0.00 \\
    \end{pmatrix*} \si{eV}\nonumber\\
    U_\mathrm{cRPA}^{\sigma\bar\sigma} &= \begin{pmatrix*}[r]
        2.28 & 1.47 & 1.45 & 1.48 & 1.87 \\
        1.47 & 2.29 & 1.78 & 1.49 & 1.54 \\
        1.45 & 1.78 & 2.54 & 1.78 & 1.51 \\
        1.48 & 1.49 & 1.78 & 2.30 & 1.55 \\
        1.87 & 1.54 & 1.51 & 1.55 & 2.51 \\
    \end{pmatrix*} \si{eV}\\
    \intertext{and their spherically symmetrized forms are}
    U_\mathrm{symm}^{\sigma\sigma} &= \begin{pmatrix*}[r]
        0.00 & 1.14 & 0.99 & 1.14 & 1.60 \\
        1.14 & 0.00 & 1.44 & 1.14 & 1.14 \\
        0.99 & 1.44 & 0.00 & 1.44 & 0.99 \\
        1.14 & 1.14 & 1.44 & 0.00 & 1.14 \\
        1.60 & 1.14 & 0.99 & 1.14 & 0.00 \\
    \end{pmatrix*} \si{eV}\nonumber\\
    U_\mathrm{symm}^{\sigma\bar\sigma} &= \begin{pmatrix*}[r]
        2.36 & 1.55 & 1.45 & 1.55 & 1.85 \\
        1.55 & 2.36 & 1.75 & 1.55 & 1.55 \\
        1.45 & 1.75 & 2.36 & 1.75 & 1.45 \\
        1.55 & 1.55 & 1.75 & 2.36 & 1.55 \\
        1.85 & 1.55 & 1.45 & 1.55 & 2.36 \\
    \end{pmatrix*} \si{eV}.
\end{align}
The order of the orbitals in these matrices is $xy$, $yz$, $z^2$, $xz$, and $x^2-y^2$.

It can be seen that there is a noticeable difference of around \SIrange{.21}{.26}{eV} between the intra-orbital terms (on the diagonal of $U^{\sigma\bar\sigma}_\mathrm{cRPA}$) corresponding to the $t_{2g}$-like orbitals and that of the $e_g$-like orbitals. However, this orbital dependence of the intra-orbital interaction is weaker than what has been found for other materials~\cite{vaugier_hubbard_2012, aichhorn_dynamical_2009}. The deviations of the various (off-diagonal) inter-orbital terms from the symmetrized form are all smaller than \SI{.1}{eV} and even slightly smaller for the same-spin case.

In the localized basis, the corresponding interaction matrices are
\begin{align}
    U_\mathrm{cRPA}^{\sigma\sigma} &= \begin{pmatrix*}[r]
        0.00 & 0.26 & -0.06 & 0.26 & 0.93 \\
        0.26 & 0.00 & 0.69 & 0.26 & 0.19 \\
        -0.06 & 0.69 & 0.00 & 0.69 & -0.12 \\
        0.26 & 0.26 & 0.69 & 0.00 & 0.19 \\
        0.93 & 0.19 & -0.12 & 0.19 & 0.00 \\
    \end{pmatrix*} \si{eV}\nonumber\\
    U_\mathrm{cRPA}^{\sigma\bar\sigma} &= \begin{pmatrix*}[r]
        \hphantom{-} 2.30 & 0.93 & \hphantom{-} 0.73 & 0.93 & \hphantom{-} 1.39 \\
        0.93 & 2.30 & 1.22 & 0.94 & 0.89 \\
        0.73 & 1.22 & 2.33 & 1.22 & 0.69 \\
        0.93 & 0.94 & 1.22 & 2.29 & 0.89 \\
        1.39 & 0.89 & 0.69 & 0.89 & 2.32 \\
    \end{pmatrix*} \si{eV}\\
    \intertext{and}
    U_\mathrm{symm}^{\sigma\sigma} &= \begin{pmatrix*}[r]
        0.00 & 0.21 & -0.04 & 0.21 & 0.94 \\
        0.21 & 0.00 & 0.69 & 0.21 & 0.21 \\
        -0.04 & 0.69 & 0.00 & 0.69 & -0.04 \\
        0.21 & 0.21 & 0.69 & 0.00 & 0.21 \\
        0.94 & 0.21 & -0.04 & 0.21 & 0.00 \\
    \end{pmatrix*} \si{eV}\nonumber\\
    U_\mathrm{symm}^{\sigma\bar\sigma} &= \begin{pmatrix*}[r]
        \hphantom{-} 2.30 & 0.90 & \hphantom{-} 0.74 & 0.90 & \hphantom{-} 1.39 \\
        0.90 & 2.30 & 1.23 & 0.90 & 0.90 \\
        0.74 & 1.23 & 2.30 & 1.23 & 0.74 \\
        0.90 & 0.90 & 1.23 & 2.30 & 0.90 \\
        1.39 & 0.90 & 0.74 & 0.90 & 2.30 \\
    \end{pmatrix*} \si{eV}.
\end{align}
Here, there is almost no orbital dependence of the intra-orbital entries in the cRPA matrices, and also the deviations of the inter-orbital matrix elements from their symmetrized form is rather small.
Notably, the interaction for parallel spins has some relatively small, but negative inter-orbital entries, both before and after symmetrization. 
This could be a sign of overscreening, which is discussed further below.

To better quantify the deviations between the unsymmetrized and symmetrized two-index matrices, we determine the maximum absolute deviation of a single matrix element, $\Delta_\mathrm{max}$, and we also compute the Frobenius norm
\begin{align}
    \Delta_\mathrm{Frob} &= \left( \sum_{mm'} \left|U^\mathrm{cRPA}_{mm'} - U^\mathrm{symm}_{mm'} \right|^2 \right)^{1/2} .
\end{align}
These deviations are summarized in \pref{tab:crpa_deviation}.
Additionally, we also quantify the deviation between the cRPA values and an alternative spherically symmetrized interaction matrix, which is obtained by first calculating parameters $U$ and $J$ according to \pref{eq:avg_umatrix} and then assuming $F_4/F_2 = 0.63$ to reconstruct the interaction matrix, as done in many DFT+$U$ and DFT+DMFT implementations.

\begin{table}
\centering
\caption{Deviations in eV between the cRPA matrices and the spherically symmetrized two-index interaction matrices for same-spin ($\sigma\sigma$) and opposite spin ($\sigma\bar\sigma$), both for the frontier and the localized basis. The symmetrization is performed either without or with the additional assumption $F_4/F_2 = 0.63$. }
\label{tab:crpa_deviation}
\begin{ruledtabular}
\begin{tabular}{lc|cc|cc}
 &  & \multicolumn{2}{c|}{$\Delta_\mathrm{max}$} & \multicolumn{2}{c}{$\Delta_\mathrm{Frob}$} \\
Basis & $F_4/F_2 = 0.63$ & $\sigma\sigma$ & $\sigma\bar\sigma$ & $\sigma\sigma$ & $\sigma\bar\sigma$ \\\hline
\multirow[t]{2}{*}{Frontier} & no & 0.05 & 0.17 & 0.12 & 0.33 \\
 & yes & 0.07 & 0.17 & 0.18 & 0.33 \\
\multirow[t]{2}{*}{Localized} & no & 0.08 & 0.05 & 0.18 & 0.11 \\
 & yes & 0.18 & 0.12 & 0.42 & 0.28 \\
\end{tabular}
\end{ruledtabular}
\end{table}

For the frontier basis, both $\Delta_\mathrm{max}$ and $\Delta_\mathrm{Frob}$ are noticeably larger for the opposite-spin interaction than for the same-spin case, due to the larger variations of the intra-orbital entries, as already discussed above. In the localized basis, the deviations from spherical symmetry are less pronounced for the opposite-spin interaction, and even slightly smaller than for the same-spin case.
On the other hand, assuming $F_4/F_2 = 0.63$ in the symmetrization has a stronger effect in the localized basis, while in the frontier basis this barely increases the overall deviation from spherical symmetry. The effect of the $F_4/F_2$ ratio discussed further in \pref{sec:effect_ratio_al}.

Compared to the interaction matrices reported in Ref.~\onlinecite{aichhorn_dynamical_2009} for LaFeAsO, the deviations from spherical symmetry are less pronounced in \CFO. 
This could be due to the different structure or coordination of the Fe cation in both systems (octahedral in \CFO versus tetrahedral in LaFeAsO), or also due to different degrees of covalency in the two systems.  
All in all, a spherical approximation of the local interaction appears to be relatively unproblematic for \CFO.

\begin{table*}
\centering
\caption{Screened and unscreened interaction parameters obtained within cRPA for the two different basis sets. The orbitally averaged nearest-neighbor Fe-Fe intersite interactions $V_\mathrm{NN}$ and $J_\mathrm{NN}$ are also listed.
}
\label{tab:crpa_results}
\begin{ruledtabular}
\begin{tabular}{ll|rrrrr|rr}
\multicolumn{2}{l|}{} & $F_0 \equiv U$ (eV) & $F_2$ (eV) & $F_4$ (eV) & $(F_2+F_4)/14 \equiv J$ (eV) & $F_4/F_2$ & $V_\mathrm{NN}$ (eV) & $J_\mathrm{NN}$ (eV) \\\hline
\multirow[t]{3}{*}{Frontier basis} & screened & 1.75 & 4.25 & 3.21 & 0.53 & 0.76 & 0.59 & 0.01 \\
 & unscreened & 14.82 & 5.91 & 3.60 & 0.68 & 0.61 & 3.83 & 0.02 \\
 & ratio (\%) & 12 & 72 & 89 & 78 & -- & 16 & 33 \\\hline
\multirow[t]{3}{*}{Localized basis} & screened & 1.25 & 7.16 & 5.73 & 0.92 & 0.80 & 0.03 & 0.00 \\
 & unscreened & 23.34 & 10.57 & 6.21 & 1.20 & 0.59 & 3.87 & 0.00 \\
 & ratio (\%) & 5 & 68 & 92 & 77 & -- & 1 & 47 \\
\end{tabular}
\end{ruledtabular}
\end{table*}

Next, we discuss the effective parameters obtained for the screened and unscreened interaction, which are listed in \pref{tab:crpa_results}.
We focus on the Hubbard $U$ and Hund $J$ obtained directly by averaging according to \pref{eq:avg_umatrix} and start the discussion with the values for the frontier basis.
The comparison of screened and unscreened values shows that $U$ is strongly screened to $U = \SI{1.75}{eV}$, which corresponds to only \SI{12}{\%} of its unscreened value, while $J$, as expected, is less screened to $J = \SI{.53}{eV}$, i.e., \SI{78}{\%} of its unscreened value.
The rather strong screening of $U$ in \CFO is indeed comparable to what has been found in cRPA calculations for rare-earth nickelates~\cite{seth_renormalization_2017, hampel_energetics_2019}, and can be explained by the close vicinity between the correlated bands and the screening bands, which even overlap slightly [see \pref{fig:bands}(b)]. 

A systematic comparison with literature data is hampered by the use of different types of correlated subspaces (full $d$-shell vs. \ttg- or \eg-only models, frontier vs. localized orbitals) or different parametrizations of the interaction Hamiltonian. The values obtained in Ref.~\onlinecite{hampel_energetics_2019} for LuNiO$_3$, if transformed from the $e_g$-only Kanamori parametrization used in \cite{hampel_energetics_2019} to the Slater parametrization used here, correspond to $U = \SI{1.3}{eV}$ and $J = \SI{0.5}{eV}$, i.e., an even smaller $U$ than in \CFO and a fairly similar $J$. Note that, due to the use of an $e_g$-only correlated subspace in \cite{hampel_energetics_2019}, the $t_{2g}$ bands will also contribute to the screening.
In cRPA calculations for other transition-metal oxides, $U$ is typically reduced to around 15-25\,\% of its bare value~\cite{vaugier_hubbard_2012, kim_strain-induced_2018, hampel_correlation-induced_2021}.
The overall magnitude of the screened $U$ and $J$ in \CFO is also comparable to values obtained for Fe in LaFeAsO, with $J = \SI{.59}{eV}$, almost equal to \CFO, and a slightly larger $U = \SI{2.14}{eV}$ \cite{aichhorn_dynamical_2009}.

In the localized basis, the unscreened values of $U$ and $J$ increase significantly compared to the frontier basis, which reflects the higher Coulomb repulsion due to the stronger localization of the corresponding orbitals. However, and perhaps surprisingly, the screened value of $U = \SI{1.25}{eV}$ is noticeably smaller than for the frontier basis and corresponds to only \SI{5}{\%} of the unscreened value.
In contrast, the screening of $J$ in the localized basis is similar (in percent) to that in the frontier basis, resulting in a screened $J = \SI{.92}{eV}$, i.e., larger than in the frontier basis.

The extremely strong screening in the localized basis in \CFO can be understood from the strong band entanglement and hybridization. As a result, virtually every band inside the energy window has a mixed character, with contributions from both the correlated subspace and the screening subspace. This results in many potential screening channels close to the Fermi energy.
To avoid this, some previous studies have used a screening subspace similar to that for the frontier orbitals also for the localized basis \cite{miyake_d-_2008, aichhorn_dynamical_2009, vaugier_hubbard_2012}. While this reduces the screening on $U$, it also introduces an inconsistency between correlated and screening subspace, and thus we have not used this approach in our work.
A small decrease in the screened $U$ (and an increase in the screened $J$) in a localized basis compared to a frontier basis has also been observed for SrMoO$_3$~\cite{hampel_correlation-induced_2021}. However, in that case, the effect is much less extreme than here. 
Generally, cRPA is known to have a tendency to overscreen, which becomes more pronounced the less the correlated and screening subspaces are separated \cite{honerkamp_limitations_2018, van_loon_random_2021}.  Consequently, one can expect that the overscreening affects both basis sets, but is particularly strong in the localized basis.
Finally, we note that our cRPA calculations are based on the low-spin state obtained within bare DFT, which might also influence the degree of screening.

Since the localized basis is similar to the correlated subspace used in our DFT+$U$ calculations, the corresponding screened interaction parameters can also be compared to the values of $U = \SI{4}{eV}$ and $J=\SI{1}{eV}$ used for our structural relaxations in \pref{sec:res_dft_relax}. Using the cRPA value of $U=\SI{1.25}{eV}$ in these calculations would lead to a rather poor description of the \Pton phase with a low-spin state of the Fe cations and essentially zero \BM mode, as shown in \pref{fig:uscan}.
On the other hand, the cRPA value of $J=\SI{0.92}{eV}$ is very close to our initial choice of $J=\SI{1}{eV}$ and (in combination with an appropriate $U$ value) appears quite reasonable. We further discuss this partial mismatch between the interaction parameters obtained from cRPA and the values required in DFT+$U$ for a good description of the charge-disproportionated state in \CFO in \pref{sec:cfo_discussion}.

Additionally, we compute the orbitally averaged nearest-neighbor intersite interaction $V_\mathrm{NN}$ and $J_\mathrm{NN}$, also shown in \pref{tab:crpa_results}.
$V_\mathrm{NN}$ assumes a moderate value of \SI{.59}{eV} in the frontier basis, screened to \SI{16}{\%} of its bare value, and comparable to LuNiO$_3$ \cite{seth_renormalization_2017} and LaFeAsO \cite{aichhorn_dynamical_2009}.
In the localized basis, the unscreened $V_\mathrm{NN}$ is comparable to that of the frontier basis, but is then completely screened, to only \SI{1}{\%} of the bare value, and thus becomes negligibly small. 
However, it is unclear whether this indeed indicates a highly efficient inter-site screening, and thus a good quality of the locality assumption for the screened interaction, or whether this values is affected by a potential overscreening.
$J_\mathrm{NN}$ is negligible in all cases (both screened and unscreened).

Finally, we also compute screened interaction parameters in the frontier basis for the relaxed structure with \Pton symmetry, and obtain only a negligible difference of \SI{3}{meV} for $U$ and less than $\SI{1}{meV}$ for $J$ compared to the $Pbnm$ structure.

\subsubsection{$F_4/F_2$ ratio in cRPA}
\label{sec:effect_ratio_al}

Next, we analyze the ratio $F_4/F_2$ obtained from the spherically averaged cRPA interaction parameters. Previous work has indicated that this ratio can deviate significantly from the value of 0.63 typically used to compute the three Slater parameters from $U$ and $J$, and can reach up to \num{.86}~\cite{vaugier_hubbard_2012}.  As shown in \pref{tab:crpa_results}, this also applies to the case of \CFO. 
The unscreened $F_4/F_2$ ratios are slightly smaller than the ``atomic'' value of 0.63, in particular for the localized basis, while in the screened case this ratio is increased to 0.80 and 0.76 for the localized and frontier basis, respectively.
Thus, while a constant ratio of $F_4/F_2 = 0.63$ still appears as a reasonable approximation for the unscreened interaction, this is no longer guaranteed in the presence of strong screening.

To assess whether a variation in $F_4/F_2$ can also affect physical properties, we estimate the energy difference between the low-spin (LS) and high-spin (HS) states of the Fe$^{4+}$ or other $d^4$ cations in the zero band-width limit by evaluating the difference in the local interaction and crystal field energies for a $t_{2g}^4 e_g^0$ and a $t_{2g}^3 e_g^1$ configuration (see also the Appendix of Ref. \onlinecite{merkel_charge_2021})
\begin{align}
E^\text{LS}-E^\text{HS} & = \frac{6}{49} F_2 + \frac{145}{441} F_4 - \Delta_\text{CF} \\[2mm] \nonumber
    & \approx \begin{cases}
        2.83 J - \Delta_\text{CF} \ \text{for}\ F_4/F_2=0.63 \\
        3.00 J - \Delta_\text{CF} \ \text{for}\  F_4/F_2=0.80
    \end{cases}
\end{align}
where $\Delta_\text{CF}$ is the local crystal-field splitting between $e_g$ and $t_{2g}$ states.
Thus, for the standard value of $F_4/F_2=0.63$, a slightly larger $J$ is necessary to obtain a high spin state. For a realistic crystal-field splitting $\Delta_\text{CF} = \SI{2.5}{eV}$, this corresponds to a difference of about \SI{0.05}{eV} in the critical $J$. While this difference does not appear very large, it could be important for systems that are very close to a transition between the high-spin and low-spin states.

\subsubsection{Different methods for disentangling the correlated and screening subspaces}
\label{sec:res_crpa_disent}

Finally, as discussed in \pref{sec:methods_h_int}, for band-structures where the correlated bands are entangled with the rest, different methods for calculating the corresponding cRPA polarization functions have been suggested. 
Here, we compare the interaction parameters obtained from the ``projector method'' \cite{kaltak_merging_2015}, which is the default method in this paper, the ``weighted method'' \cite{sasioglu_effective_2011}, and the ``disentanglement method'' \cite{miyake_ab_2009}, as implemented in VASP. The comparison of different interaction parameters is shown in \pref{tab:crpa_results_projectors}. 

\begin{table}
\centering
\caption{Screened interaction parameters obtained with different band disentanglement methods for the two different basis sets. We compare the parameters $U$ and $J$, the ratio $F_4/F_2$, and the averaged intersite interaction $V_\mathrm{NN}$.
}
\label{tab:crpa_results_projectors}
\begin{ruledtabular}
\begin{tabular}{ll|rrrr}
\multicolumn{2}{l|}{Interaction params.} & $U$ (eV) & $J$ (eV) & $F_4/F_2$ & $V_\mathrm{NN}$ (eV) \\\hline
\multirow[t]{3}{*}{Frontier} & proj. & 1.75 & 0.53 & 0.76 & 0.59 \\
& weighted & 1.74 & 0.53 & 0.76 & 0.60 \\
& disent. & 3.12 & 0.55 & 0.73 & 1.35 \\\hline
\multirow[t]{3}{*}{Localized} & proj. & 1.25 & 0.92 & 0.80 & 0.03 \\
& weighted & 0.46 & 0.80 & 0.98 & 0.03 \\
& disent. & 5.41 & 1.07 & 0.64 & -0.07 \\
\end{tabular}
\end{ruledtabular}
\end{table}

In the frontier basis, the projection and weighted methods give almost identical results for all interaction parameters, while the disentanglement method predicts larger values, and thus less screening, especially for the on- and intersite Coulomb repulsion $U$ and $V_\mathrm{NN}$.
The former two methods likely agree well because the Fe-dominated bands captured by the frontier basis only slightly overlap with the O-dominated bands. This in turn means that almost all bands clearly belong to either the correlated or uncorrelated subspace and only a few bands right below the frozen window at \SI{-1.1}{eV} partially belong to both subspaces. Furthermore, both methods work on the unaltered DFT band-structure.
In contrast, neglecting the hybridization between correlated and uncorrelated subspaces, will slightly affect the band structure, in particular in the overlapping region below \SI{-1}{eV}, which can explain the more significant deviation of the disentanglement method compared to the other two method.

In the localized basis, the differences in interaction parameters, in particular $U$, are much more pronounced between the different methods. Again, the disentanglement method results in a significantly higher value of $U$ and thus less screening. However, also the projection and weighted methods now differ quite drastically, with a very small $U = \SI{0.46}{eV}$ obtained from the latter.
The large spread of the calculated interaction parameters for the localized basis is most likely a result of the strong band entanglement, such that small differences in the way the two subspaces are separated can lead to rather large differences in the corresponding polarization functions.

On first view, the interaction parameters obtained from the disentanglement method appear quite reasonable and compatible with the values used in our DFT+$U$ calculations in \pref{sec:res_dft_relax}. However, ignoring the hybridization between correlated and uncorrelated subspaces in the localized basis reduces the band width of the correlated bands to only  \SI{.6}{eV} for the Fe-\eg and \SI{.5}{eV} for the Fe-\ttg bands, and thus leads to severe changes compared to the DFT bands. This is very different from the situation for which this method was introduced \cite{miyake_ab_2009}.
Therefore, this match of the interaction values seems likely to be a coincidence given the strong modification of the disentangled band structure.

The intersite terms are always negligibly small for the localized basis.

\section{Summary and discussion}
\label{sec:cfo_discussion}

In this work, we have presented cRPA calculations of the screened interaction parameters for the charge-disproportionated insulator \CFO. 
Thereby, we have compared results for two different types of Wannier functions used to describe the Fe-$d$ states. First, a frontier orbital basis, incorporating the hybridization with the oxygen ligands, which describes only the bands immediately around the Fermi level and maps well on simple models involving only Fe cations with formal charge states. Second, a more localized basis resembling an atomic orbital-like basis, similar to the basis typically used in DFT+$U$ calculations.

We have shown that cRPA predicts \CFO to have a strongly screened Hubbard $U$ and a relatively large Hund's interaction $J$, consistent with other charge-disproportionated insulators such as the rare-earth nickelates~\cite{seth_renormalization_2017, hampel_energetics_2019}. This supports the picture that the basic physics of these systems can be described within a minimal model where the insulating state is driven by the Hund's interaction rather than a strong Hubbard $U$~\cite{strand_valence-skipping_2014, subedi_low-energy_2015, isidori_charge_2019, merkel_charge_2021}.
Additionally, the DFT band structure indicates that \CFO has a relatively small but positive charge-transfer energy.

We have also tested the quality of certain approximations typically applied to parametrize the interaction matrices in first-principles calculations, such as the assumption of spherical symmetry and a fixed ratio between the Slater integrals $F_2$ and $F_4$. Indeed, the spherical approximation appears relatively uncritical, both for the more localized and the frontier basis, with the largest deviations from spherical symmetry observed for the intra-orbital interactions in the frontier basis, which vary by around \SIrange{0.2}{0.25}{eV}.
The calculated $F_4/F_2$ ratio was found to deviate noticeably from the commonly used fixed ratio of 0.63, in agreement with previous cRPA calculations for other transition-metal oxides~\cite{vaugier_hubbard_2012}. 
This can potentially become important when simulating a material close to a transition, e.g., between the high- and low-spin states.

In addition, we have discussed the choice of interaction parameters in DFT+$U$ calculations, to obtain structural parameters for both the high- and low-temperature structures that are in good agreement with experimental data.
We have shown that a simple A-type antiferromagnetic order and moderate electronic interactions suffice to reproduce the experimentally observed low-temperature structure. By subsequently neglecting the local electron-electron interactions and magnetic order, a good description of the experimental high-temperature structure can also be obtained.

However, our study also highlights some 
difficulties of directly using interaction parameters obtained from cRPA in Hubbard-corrected DFT-based methods such as DFT+$U$ or DFT+DMFT, in particular for systems with strong entanglement between the correlated and the uncorrelated subspaces. The cRPA values obtained for $U$ in the localized basis are clearly too small to obtain a good description of the charge-disproportionated state of \CFO in DFT+$U$ calculations.
While this apparent mismatch can, at least partially, be attributed to the well-established tendency for overscreening within the random-phase approximation \cite{honerkamp_limitations_2018, van_loon_random_2021}, in the present case it is also, to a large extent, a result of the difficulty to disentangle the electronic bands between the correlated and uncorrelated subspaces.
From our comparison between different methods in \pref{sec:res_crpa_disent} it becomes clear that, in situations with strong entanglement, the results can vary widely and none of the methods is likely to produce interaction parameters that can reliably be used in first-principles calculations.
Furthermore, screening is fundamentally a frequency-dependent phenomenon, while DFT+$U$ (and also standard DFT+DMFT) uses constant, frequency-independent interaction parameters. It thus remains unclear whether using the zero-frequency value of the screened interaction is necessarily the best choice to be used in such calculations (see also Ref.~\onlinecite{casula_low-energy_2012}).
The further simplifications within the DFT+$U$ method, i.e., the Hartree-Fock-like approximation for the local interaction and the treatment of intersite effects only within the standard semi-local DFT approximations, can also affect the optimal choice for the effective interaction parameters.
Finally, it would be desirable to perform cRPA calculations for \CFO also in the high-spin state, either by constraining the orbital occupations to a $t_{2g}^3 e_g^1$ configuration or by performing cRPA calculations for a magnetically ordered state. 

While a solution of these issues goes beyond the scope of the present work, our results can serve as starting point for future attempts aimed at improving the currently available methods for calculating effective interaction parameters. 

\section*{Data Availability}

The supporting data for this article are openly available from the Materials Cloud Archive \cite{materialscloud_crpa_cfo}.

\begin{acknowledgments}
We thank Alexander Hampel and Alexander Gillmann, who performed initial calculations related to this project.
This research was supported by ETH Zurich and a grant from the Swiss National Supercomputing Centre (CSCS) under project ID s1128.
\end{acknowledgments}

\appendix*

\section{Equations for the spherically symmetric interaction}

Here, we summarize properties related to the tensors $\alpha^{(k)}$ relevant for \pref{sec:method_crpa_avg}. 
The $\alpha^{(k)}$ tensors are orthogonal with the normalization
\begin{align}
    &\sum_{mm'm''m'''} \alpha^{(k_1)}_{mm'm''m'''} \alpha^{(k_2)}_{mm'm''m'''} = c_{lk_1} \delta_{k_1,k_2} \nonumber\\
    &\text{with}\quad
    c_{lk} := \frac{(2l+1)^4}{2k+1} \begin{pmatrix} l & k & l \\ 0 & 0 & 0 \end{pmatrix}^4 ,
    \label{eq:ortho_alpha}
\end{align}
which directly allows us to derive \pref{eq:avg_umatrix_slater}.
From \pref{eq:avg_umatrix_slater} and \pref{eq:avg_umatrix}, a general expression for $J$ for all $l$ follows as
\begin{align}
    J = \frac{2l+1}{2l} \sum_{k \geq 2} F_k
    \begin{pmatrix} l & k & l \\ 0 & 0 & 0 \end{pmatrix}^2 ,
    \label{eq:general_j_slater}
\end{align}
which gives $J = (F_2 + F_4)/14$ for $l=2$.

\bibliography{bibfile}

\end{document}